\documentstyle[preprint,aps]{revtex}
 \begin{document}
\preprint{USACH/99/11}
\title{FERMIONS ON NON-TRIVIAL TOPOLOGIES}
\author{ J. Gamboa\thanks{E-mail: jgamboa@lauca.usach.cl}}
\address{Departamento de F\'{\i}sica, Universidad de Santiago de Chile,
Casilla 307, Santiago 2, Chile }

\maketitle

\begin{abstract}
An exact expression for the Green function of a purely fermionic
system moving on the manifold $\Re \times \Sigma^{D-1}$, where
$\Sigma^{D-1}$ is a $(D-1)$-torus, is found. This expression
involves the bosonic analog of $\chi_n = e^{in\theta}$
corresponding to the irreducible representation for the n-th class
of homotopy and in the fermionic case for $D=2$ and $3$, $\chi_n$
is a measure of the statistics of the particles. For higher
dimensions ($D \geq 4$), there is no analogue interpretation however 
this could, presumably, indicate a generation of mass as in quantum field
theories at finite temperature.
\end{abstract}

The scattering of fermions by vortices is a problem that has
important applications in particle physics\cite{alford} and in the
realistic description of the Aharonov-Bohm effect\cite{ab}.
Although this topic was partially discussed in the
literature and new corrections to the AB cross section have been  
found \cite{varios}, as far as we know, there are no studies
trying to understand the connection between these processes and
the possibility of summing over the homotopy classes as in the bosonic
case\cite{Schulman},\cite{witt}.

If we have a multiply connected manifold with one vortex, the propagation amplitude for a spinless particle is

\begin{equation}
G[x,x^{'}] = \sum_n \chi_n G_n [x, x^{'}], \label{1}
\end{equation}
where $G_n [x, x^{'}]$ is the propagation amplitude for a given winding number $n$ and $\chi_n$
is an irreducible representation for the n-th homotopy class. The composition law for
the Green function implies that $\chi_n = e^{in\delta}$ , where $\delta$ is a quantum phase which in the case of the Aharonov-Bohm effect is just the magnetic flux. Equation (\ref{1}) was first conjectured by Schulman in \cite{Schulman} and proved rigourosly by Laidlaw and C. de Witt in \cite{witt}.

Several questions follow from this result: What is the analog of (1) for particles
with spin?, What is (if it exists) the physical meaning of the quantum phase for fermions?.

The purpose of this paper is to give an answer to the above
questions and to extend and sketch possible applications for more
complicated topologies.

In order to expose our work more systematically, let us
start noticing that the Dirac propagator can be written as

\begin{eqnarray}
G_D[x,x^{'}] &=& \biggl< \frac{1}{{\partial \hspace{-.6em} \slash \hspace{.15em}}}\biggr>,
\nonumber
\\
&=&\int_0^\infty dT T^{-\frac{D}{2}} \biggl( \frac{\gamma^\mu \Delta x_\mu}{T} \biggr)
e^{-\frac{{(\Delta x)}^2}{2T} },\label{2}
\\
&=& \int \frac{d^Dp}{{(2\pi)}^D} \frac{({p \hspace{-0.6em} \slash \hspace{0.15em}} )}{p^2 }
e^{ip.\Delta x}.
 \label{3}
\end{eqnarray}

Although in this calculation one is inverting the first order operator
$(i{\partial \hspace{-0.6em} \slash \hspace{0.15em}} )$, and therefore one needs one boundary condition, the relation

\begin{equation}
G_D[x,x^{'}] = (i {\partial \hspace{-.6em} \slash \hspace{.15em}} ) \, \Delta ( x,x^{'}),
\label{4}
\end{equation}
where $\Delta ( x,x^{'})$ is the propagator for a spinless particle defined as

\begin{equation}
\Delta ( x,x^{'}) = \int \frac{d^Dp}{{(2\pi)}^D} \frac{e^{ip.\Delta x}}{p^2 }, \label{5}
\end{equation}
alert us that the boundary condition

\begin{equation}
x^\mu (\tau_1) = x^\mu_1, \,\,\,\,\,\,\,\,\,\,\,\,\,\,\,\,\,\,\,\,\,\,
x^\mu (\tau_2) = x^\mu_2, \label{6}
\end{equation}
has been used with $\tau$ being the proper-time.

Keeping in mind the above comments, let us start by considering the simplest fermionic system,
namely, a free massless \lq toy fermion\rq$\,\,$  living on a $0+1$-dimensional space\footnote{Although this is a non physical system, will be useful for the generalization to higher dimensions.}. 
The topology for this
system is the line $\Re^1$ and the boundary conditions (\ref{6}) become 

\begin{equation}
x(\tau_1) = x_1, \,\,\,\,\,\,\,\,\,\,\,\,\,\,\,\,\,\,\,\,\,\,
x (\tau_2) = x_2, \label{66}
\end{equation}

Using (\ref{2}) one finds that the propagator is

\begin{eqnarray}
G_D[x,x^{'}] &=& \lim_{m\rightarrow 0} \biggl<\frac{1}{\partial +m}\biggr> \nonumber
\\
&=& \lim_{m\rightarrow 0} \int_0^\infty dT T^{-\frac{1}{2}} ( \frac{\Delta x}{T} - m) \,\,
e^{-\frac{{(\Delta x)}^2}{2T} - \frac{m^2T}{2}}, \label{7}
\end{eqnarray}
where in (\ref{7}) an infrared regulator $m$ has been introduced in order to make the
operator ${( \partial + m)}^{-1}$ well defined.

The propagator (\ref{7}) can be evaluated by means of the identity\cite{grad}

\begin{equation}
\int_0^\infty dx x^{\nu -1} \,\,e^{ -\frac{\beta}{x} - \gamma x} = 2
{\biggl[ \frac{\beta}{\gamma}\biggr]}^{\frac{\nu}{2}}\,\, K_\nu ( 2 \sqrt{\beta \gamma}),
\label{gr}
\end{equation}
and it becomes

\begin{eqnarray}
G_D[x,x^{'}] &=& \lim_{m\rightarrow 0} \Delta x \int_0^\infty dT T^{-\frac{3}{2}} \,\,
e^{-\frac{{(\Delta x)}^2}{2T} - \frac{m^2T}{2}} -
\lim_{m\rightarrow 0} \,\,m\,\,\int_0^\infty dT T^{-\frac{1}{2}} \,\,
e^{-\frac{{(\Delta x)}^2}{2T} - \frac{m^2T}{2}}, \nonumber
\\
&=& \lim_{m\rightarrow 0} \Delta x {\biggl[ \frac{{(\Delta x)}^2}{m^2}\biggr]}^{-\frac{1}{4}}\,\,
 K_{-\frac{1}{2}} ( m \sqrt{{(\Delta x)}^2}, \label{8}
\end{eqnarray}
where the last term in the RHS in (\ref{8}) have been removed because it does not contribute
in the limit $m\rightarrow 0$. Using $K_\nu (z) \sim x^\nu$ for $x<<1$, one finds after
taking the limit

\begin{equation}
G_D[x,x^{'}] = \frac{\Delta x}{2 \vert \Delta x\vert} = \frac{1}{2}
sgn ( x - x^{'}), \label{10}
\end{equation}
and, as a consequence, the propagation amplitude becomes the sign function and the relation

\begin{equation}
\partial G_D[x,x^{'}] = \delta (x, x^{'}), \label{11}
\end{equation}
is correctly satisfied.

The next step will be to complicate a little bit the problem considering the $S^1$ topology instead of $\Re^1$. In this case the boundary conditions (\ref{6}) must be
modified as follow

\begin{equation}
x (\tau_1) = x_1, \,\,\,\,\,\,\,\,\,\,\,\,\,\,\,x(\tau_2) = x_2 + 2n\pi, \label{12}
\end{equation}
where $n=0,\pm 1, \pm 2,  ...$, is the winding number around the circle and in
equation (\ref{10}) one should make the replacement $\Delta x \rightarrow \Delta x + 2n\pi$ and (\ref{10}) becomes

\begin{equation}
G_{Dn}[x,x^{'}] =  \frac{1}{2} sgn ( x - x^{'} + 2n\pi). \label{13}
\end{equation}

Then, naively, (\ref{13}) would be the Green function for the n-th class of homotopy. The total Green function is obtained by summing over $n$ and weighting each contribution with a factor $\chi_n$, {\it i.e.} one should replace (\ref{13}) by

\begin{equation}
G_D[x,x^{'}] =  \sum_n \chi_n \, G_{Dn}[x,x^{'}]. \label{14}
\end{equation}

Since (\ref{14}) must satisfy the composition law,
\begin{equation}
G[x,x^{\prime}] = \int dy \,\,G[x,y]\,\,G[y,x^{\prime}], \label{com}
\end{equation}
then (\ref{14}) and (\ref{com}) imply
\begin{equation}
\chi_n = e^{in\theta}. \label{factor}
\end{equation}

The next question is to identify the interval where $x$ is defined. One can give an answer to this question writing firstly, the Green function $G_D(x)$ as 
\begin{eqnarray}
G_D(x) &=& \sum_{n=-\infty}^\infty e^{in\theta} G_{Dn} (x), \nonumber
\\
&=& G_{D0}(x) + \sum_{n=1}^\infty e^{in\theta} G_{Dn}(x) + \sum_{n= -\infty}^{-1} e^{in\theta} G_{Dn} (x), \label{corr}
\end{eqnarray}
where $G_{Dn} (x)$ is defined in (\ref{13}).

If $x$ is defined over the interval $0<x<2\pi$, then the three terms in the RHS of (\ref{corr}) are $1, \sum_{n=1}^\infty e^{in\theta}$ and $\sum_{n=1}^\infty e^{-in\theta}$ respectively, and $G_{D} (x)$  defined in  (\ref{corr}) does not satisfy (\ref{11}). However if $x$ is defined over $-\pi<x<\pi$, then (\ref{11}) is satisfied because the zero mode in (\ref{corr}) is just the sign function.

Thus, on this interval the function 
\begin{equation}
G_D[x,x^{'}] =  \frac{1}{2} \sum_n e^{in\theta}sgn ( x - x^{'}+ 2n\pi), \label{166}
\end{equation}
is the correct Green function for a one-dimensional \lq fermion\rq $\,\,$moving on $S^1$.

Let us now analyse the physical meaning of $\theta$. Under the shift $x^{'} \rightarrow x^{'} + 2\pi$
the Green function (\ref{166}) trasforms as 

\begin{equation}
G_D[x,x^{'} + 2\pi] =  \frac{1}{2} \sum_{n=-\infty}^{\infty}\, e^{in\theta}\,\,sgn
[ x - x^{'}+ 2(n-1)\pi], \label{17}
\end{equation}
and redefining $n-1=m$, one finds

\begin{equation}
G_D[x,x^{'}+ 2\pi] = e^{i\theta} \,\,G_D[x,x^{'}]. \label{18}
\end{equation}

Thus, if we recall that $\psi (x) = \int dx^{'} G[x,x^{'}] \psi (x^{'})$, then one obtains

\begin{equation}
\psi (x + 2\pi) = e^{-i\theta} \psi (x), \label{19}
\end{equation}
implying that after one turn, the spinor changes in a phase factor and $\theta$ becomes a measure of the \lq statistics\rq $\,\,$ however the question is; what statistics?.

This question can be answered as follows; one can consider the above example as describing the relative motion of two fermions with the center of mass decoupled. Thus, our effective one dimensional \lq fermion\rq $\,\,$ is described by the Lagrangian

\begin{equation}
L = \frac{i}{2} \psi^* {\dot \psi} + ..., \label{fe}
\end{equation}
where $...$ means terms involving the center of mass and, thus, the phase factor in (\ref{19}) becomes a measure of the
interchange  of two particles.

The above results can be easily generalized to $D$-dimensions for the $n$-torus. Indeed
let us consider a spacetime with the topology $T\times \Sigma^{D-1}$ where $T=\Re^1$ is
the time and $\Sigma^{D-1}$ is $(D-1)$ times $S^1$. Then the boundary conditions (\ref{6})
become

\begin{eqnarray}
x^0 (\tau_1) &=& x^1_1\,\,\,\, \,\,\,\,\,\,\,\,x^0 (\tau_2) = x^0_2, \nonumber\\
x^1 (\tau_1) &=& x^1_1, \,\,\,\, \,\,\,\,\,\,\,\,x^1 (\tau_2) = x^1_2 + 2n_1\pi, \nonumber\\
x^2 (\tau_1) &=&x^2_1,\,\,\,\, \,\,\,\,\,\,\,\,x^2 (\tau_2) = x^2_2 + 2n_2\pi, \label{wn}\\
\vdots \nonumber\\
x^{D-1} (\tau_1) &=& x^{D-1}_1,\,\,\,\, \,\,\,\,\,\,\,\,x^{D-1} (\tau_2) =
x^{D-1}_2 + 2n_{D-1}\pi, \nonumber
\end{eqnarray}
where $n_\beta$ are the winding numbers.

Using (\ref{wn}), the Green function for the $n$-th class of homotopy is
\begin{eqnarray}
G_{Dn}[&x_2&, x_1] = \lim_{m\rightarrow 0} \int_0^\infty dT T^{-D/2}
\biggl[ \frac{\gamma^\mu \Delta x_\mu}{T} - m\biggr]\,\,e^{- \frac{{(\Delta x)}^2}{2T} -
\frac{m^2}{2}T}, \nonumber
\\
&=&  \lim_{m\rightarrow 0} ( \gamma^0 \Delta x_0 + \sum_{\alpha = 1}^{D-1}
\gamma^\alpha \Delta x_\alpha \label{p})\int_0^\infty dT T^{-\frac{D}{2} - 1}
\exp \biggl[ -\frac{ {(\Delta x_0)}^2 + \sum_{\beta = 1}^{D-1} {(\Delta x_\beta -
2n_\beta \pi)}^2 }{2T} - \frac{m^2T}{2} \biggr]. \label{rr}
\end{eqnarray}

The integral in (\ref{rr}) can be computed using (\ref{gr}) and in the massless limit
(\ref{rr}) becomes
\begin{equation}
G_{Dn} [x_2,x_1] = \frac{\gamma^0 \Delta x_0 + \sum_{\alpha =
1}^{D-1} \gamma^{\alpha} ( \Delta x_{\alpha} - 2n_{\alpha} \pi)}
{{\vert{ \Delta x_0 + \sum_{\alpha = 1}^{D-1} ( \Delta x_{\alpha}
- 2n_{\alpha} \pi)}\vert}^{D}}, \label{li}
\end{equation}
containing (\ref{13}) as a particular case (after removing
$\Delta x_0$ and putting $\gamma^{\alpha} =1$).

>From (\ref{li}) one can conjecture the total Green function
\begin{equation}
G_D [x_2, x_1] = \sum_{n_1} \sum_{n_2} ...\sum_{n_{D-1}} \chi_{n_1 n_2 ...n_{D-1}} \,
\,G_{n_1 n_2 ...n_{D-1}}
[x_2, x_1], \label{ggf}
\end{equation}
where $\chi_{n_1 n_2...n_{D-1}}$ is the generalized $n$-th homotopy class factor.

However, by a theorem of homotopy theory\cite{sch}, that essentially corresponds to the
generalization of the composition law for the Green function, the homotopy factors must satisfy
\begin{equation}
\chi_{n_1 n_2 ...n_{D-1}} = \chi_{n_1 +n_2 +...+n_{D-1}}, \label{hom}
\end{equation}
and, as a consequence, each $\chi_{n_i}$ must be an exponential and (\ref{ggf}) becomes 
\begin{equation}
G_D [x_2, x_1] = \sum_{n_1} \sum_{n_2}...\sum_{n_{D-1}} \,e^{i(n_1 \theta_1 + n_2
\theta_2 ...+n_{D-1} \theta_{D-1})}
\, G_{n_1 n_2...n_{D-1}} [x_2, x_1], \label{final}
\end{equation}
which is the Green function for a massless fermion living on the manifold  $\Re^1\times \Sigma^{D-1}$, where $\Sigma^{D-1}$ is the $(D-1)$-torus.

Roughly speaking, the case $D=2$ is formally equivalent to (18) because for $D=2$ the
spatial manifold ${\cal M}_{spatial} \sim S^1$, and (\ref{ggf}) becomes the Green function
for a fermion on a cylinder. For $D=3$, ${\cal M}_{spatial} \sim T^2$
is a torus and the phase factor is also a measure of the (fractional) statistics as
was discussed by Einarsson in \cite{ei}. For higher dimensions ($D\geq 4$) the phases are not 
related with the statistics and, presumably this could be related with generation of mass as in
quantum field theory at finite temperature.

One could try also to include the more interesting cases such as the
Aharonov-Bohm effect with $N$ aligned solenoids, however, although
this problem is more complicated because the group of paths is non
abelian, one could extract some information from the problem
considered here. For a discussion about this problem
in the bosonic case see \cite{nambu}.

In conclusion, we have generalized the propagation amplitude for the fermionic case and extended this result to the $n$-torus case for arbitrary
dimensions. Our result could yield interesting applications in
$(2+1)$-dimensions and, particularly, a detailed study of the
fractional statistics on a torus\cite{ei} following these lines
could be relevant in applications to condensed matter physics. More
details about these applications will be published elsewhere.

I gratefully acknowledge numerous conversations with J. Alfaro, M. Ba\~nados, J. L. Cort\'es, V. O. Rivelles, M. Plyushchay and J. Zanelli. This work was partially supported by grants from FONDECYT-Chile (1980788) and DICYT-USACH. This paper was finished when the author was visiting the Department of Physics and Astronomy of the University of Kansas. I would like to thank D. W. McKay and J. P. Ralston for the hospitality and the
Fundaci\'on Andes for financial support.

\end{document}